# Predicting Human Preferences Using the Block Structure of Complex Social Networks

Roger Guimerà[1,2]*, Alejandro Llorente[3], Esteban Moro[4,5,3], Marta Sales-Pardo[2]

1 Institució Catalana de Recerca i Estudis Avançats, Barcelona, Catalonia, Spain, 2 Departament d'Enginyeria Qumica, Universitat Rovira i Virgili, Tarragona, Catalonia, Spain, 3 Instituto de Ingeniería del Conocimiento, Universidad Autónoma de Madrid, Madrid, Spain, 4 Departamento de Matemáticas & Grupo Interdisciplinar de Sistemas Complejos, Universidad Carlos III de Madrid, Leganés, Madrid, Spain, 5 Instituto de Ciencias Matemáticas, Consejo Superior de Investigaciones Científicas-Autonomous University of Madrid-Universidad Complutense de Madrid-Universidad Carlos III de Madrid, Madrid, Madrid, Spain

**Abstract**

With ever-increasing available data, predicting individuals' preferences and helping them locate the most relevant information has become a pressing need. Understanding and predicting preferences is also important from a fundamental point of view, as part of what has been called a "new" computational social science. Here, we propose a novel approach based on stochastic block models, which have been developed by sociologists as plausible models of complex networks of social interactions. Our model is in the spirit of predicting individuals' preferences based on the preferences of others but, rather than fitting a particular model, we rely on a Bayesian approach that samples over the ensemble of all possible models. We show that our approach is considerably more accurate than leading recommender algorithms, with major relative improvements between 38% and 99% over industry-level algorithms. Besides, our approach sheds light on decision-making processes by identifying groups of individuals that have consistently similar preferences, and enabling the analysis of the characteristics of those groups.

**Funding:** This work was supported by a James S. McDonnell Foundation Research Award (RG and MSP), grants PIRG-GA-2010-277166 (RG) and PIRG-GA-2010-268342 (MSP) from the European Union, and grants FIS2010-18639 (RG and MSP), FIS2006-01485 (MOSAICO) (EM) and FIS2010-22047-C05-04 (EM) from the Spanish Ministerio de Economía y Competitividad. The funders had no role in study design, data collection and analysis, decision to publish, or preparation of the manuscript.

**Competing Interests:** The authors have declared that no competing interests exist.

* E-mail: roger.guimera@urv.cat

## Introduction

Humans generate information at an unprecedented pace, with some estimates suggesting that in a year we now produce on the order of $10^{21}$ bytes of data, millions of times the amount of information in all the books ever written [1]. In this context, predicting individuals' preferences and helping them locate the most relevant information has become a pressing need. This explains the outburst, during the last years, of research on recommender systems, which aim to identify items (movies or books, for example) that are potentially interesting to a given individual [2–4].

However, understanding and ultimately predicting human preferences and behaviors is also important from a fundamental point of view. Indeed, the digital traces that we leave with all sorts of everyday activities (shopping, communicating with others, traveling) are ushering in a new kind of computational social science [5,6], which aims to shed light on human mobility [7,8], activity patterns [9], decision-making processes [10], social influence [11–13], and the impact of all these in collective human behavior [14,15].

Existing recommender systems are good at solving the practical problem of providing quick estimates of individuals' preferences, but they often emphasize computational performance over other important questions such as whether the algorithms are mathematically well-grounded or whether the implicit models and assumptions are easy to interpret (and therefore to modify and fine tune). In contrast, algorithms that are based on plausible, easily-interpretable assumptions and that are based on solid mathematical grounds are useful in themselves and, arguably, hold the most potential to advance in the solution of the problem at the fundamental and practical levels. Here we present one such approach and show that it performs better than state-of-the-art recommender systems.

In particular, we focus on what is called collaborative filtering [16], namely making predictions about preferences based on preferences previously expressed by users. The underlying assumption in virtually all collaborative filtering approaches is that similar people have similar "interactions" with similar items. This consideration is usually taken into account heuristically. For example, in memory-based methods [16], one tries to identify users that are similar to the one for which we seek a prediction; or items that are similar to the target item. From these "neighbors" one then obtains a weighted average. In matrix factorization approaches [17], one assumes that each user and item can be characterized by a low-dimensional "feature vector," and that the rating of an item by a user is the product of their feature vectors.

In contrast, we base our predictions in a family of models [18–21] that have been developed and are widely used by sociologists as plausible models of complex social networks, that is, of how social actors establish relationships (friendship relationships with each other, or membership relationships with institutions, for example). In this family of models, social actors are divided into groups and relationships between two actors are established depending solely on the groups to which they belong. Because of their simplicity and their explanatory power, these models are increasingly being studied as general models of complex (not necessarily social) networks [22–24].

In the context of predicting human preferences, block models assume that users and items can be simultaneously classified into categories, and that the category of the user and the category of the item fully determine the rating. Therefore, the model is extremely easy to interpret. Additionally, our algorithm is mathematically sound because it uses a Bayesian approach that deals rigorously with the uncertainty associated with the models that could potentially account for observed users' ratings. Indeed, our approach averages over the ensemble of *all* possible groupings of users and items, exploiting the formal analogies that exist between statistical inference and statistical physics [25].

Finally, our algorithm sheds light on the factors determining preferences because it allows one to study the groupings that have the most explanatory power or that accurately account for certain features of the users' ratings.

## Bayesian Predictions Based on the Ensemble of Stochastic Block Models

Consider the observed ratings $R^O$, whose element $r_{ui}^O$ represents the rating of user $u$ on item $i$ (Fig. 1). Note that not all elements in this "matrix" are defined, since only some pairs $(u,i)$ are actually observed; we call $O$ the set of observed $(u,i)$ pairs. Like in collaborative filtering approaches [2,3], we assume that these observations are all the information that the algorithm can use to make predictions about unobserved ratings (in other words, we do not use any information about users or items other than past ratings). Our problem is then to estimate the probability $p(r_{ui}=r|R^O)$ that the unobserved rating of item $i$ by user $u$ is $r_{ui}=r$, given the observation $R^O$.

Let's assume that the observed ratings can be explained by one of the models in a family $\mathcal{M}$ of generative models. Then,

$$p(r_{ui}=r|R^O) = \int_\mathcal{M} dM\, p(r_{ui}=r|M) p(M|R^O), \qquad (1)$$

where $p(r_{ui}=r|M)$ is the probability that $r_{ui}=r$ if the ratings where actually generated using model $M$, and $p(M|R^O)$ is the plausibility of model $M$ given the observation. Using Bayes theorem Eq. (1) becomes

$$p(r_{ui}=r|R^O) = \frac{\int_\mathcal{M} dM\, p(r_{ui}=r|M) p(R^O|M) p(M)}{\int_\mathcal{M} dM'\, p(R^O|M') p(M')}, \qquad (2)$$

where $p(R^O|M)$ is the probability that model $M$ gives rise to $R^O$ among all possible ratings (or the likelihood of the model), and $p(M)$ is the a priori probability that model $M$ is the correct one (or prior). This equation is formally equivalent to those derived in the context of network inference [22] and, more broadly, to those used in Bayesian model averaging [26].

Although Eq. (2) is the correct probabilistic treatment of $R^O$ for inference of unobserved ratings, in practice predictions will only be accurate if the models in $\mathcal{M}$ (or at least some of them) correctly describe how users actually rate items. Additionally, the models need to be simple enough that they are analytically or computationally tractable.

We consider the family $\mathcal{M}_{\text{SBM}}$ of stochastic block models [19–22]. In a stochastic block model, users and items are partitioned into groups and the probability that a user rates an item with $r_{ui}=r$ depends, exclusively, on the groups $\sigma_u$ and $\sigma_i$ to which the user and the item belong, that is

$$p(r_{ui}=r|M) = q_r(\sigma_u, \sigma_i) \in [0,1], \qquad (3)$$

with $\sum_r q_r(\sigma_u, \sigma_i) = 1$.

Consider the case in which ratings can take $K$ different values $r \in \{1, \ldots, K\}$ (we use the labels $1, \ldots, K$ for simplicity, but the only requirement is that there are $K$ non-overlapping classes, which do not need to be ordinals). Under the assumption of no prior knowledge about the models ($p(M)=\text{const.}$), one can partially integrate Eq. (2) (see Methods) to obtain

$$p_{\text{SBM}}(r_{ui}=r|R^O) =$$

$$\frac{1}{Z} \sum_{\substack{P_U \in \mathcal{P}_U \\ P_I \in \mathcal{P}_I}} \left( \frac{n^r_{\sigma_u \sigma_i} + 1}{n_{\sigma_u \sigma_i} + K} \right) e^{-\mathcal{H}(P_U, P_I)}, \qquad (4)$$

where the sum is over all possible partitions of users and items into groups ($\mathcal{P}_U$ and $\mathcal{P}_I$, respectively), $n^r_{\sigma_u \sigma_i}$ is the number of $r$-ratings observed from users in group $\sigma_u$ to items in group $\sigma_i$, and $n_{\sigma_u \sigma_i} = \sum_{k=1}^K n^k_{\sigma_u \sigma_i}$ is the total number of observed ratings from users in $\sigma_u$ to items in $\sigma_i$. The "Hamiltonian" $\mathcal{H}(P_U, P_I)$, which weights the contribution of each partition, depends only on the partition

$$\mathcal{H}(P_U, P_I) = \sum_{\alpha, \beta} \left[ \ln(n_{\alpha\beta} + K - 1)! - \sum_{k=1}^K \ln(n^k_{\alpha\beta})! \right] \qquad (5)$$

and $Z = \sum e^{-\mathcal{H}}$ is the partition function.

Although carrying out the exhaustive summation over all partitions in Eq. (4) is unfeasible, one can estimate $p_{\text{SBM}}(r_{ui}=r|R^O)$ using Metropolis sampling [22,25,27]. Given these probabilities, our prediction for a given rating is the one that maximizes the probability

$$r_{ui}^* = \arg\max_r p_{\text{SBM}}(r_{ui}=r|R^O). \qquad (6)$$

## Benchmark Algorithms

To test how accurately our stochastic block model (SBM) algorithm predicts human preferences, we compare its performance to that of some of the most accurate algorithms in the literature of collaborative filtering recommender systems (see Methods for details) [4]. First, we consider a matrix factorization method [17] based on singular value decomposition (SVD) [28], which uses stochastic gradient descent to minimize the deviations between model predictions and observed ratings [17]. We use two implementations of this algorithm: our own implementation (SVD1) as well as a highly optimized implementation provided by LensKit framework (SVD2) [4]. Second, we consider an algorithm based on the similarity between items [4,29], and again use the LensKit implementation (Item-Item). Additionally, we consider a baseline naive recommender, where the rating of an item by a user is simply the average rating of the item by all users that have rated it before [4].

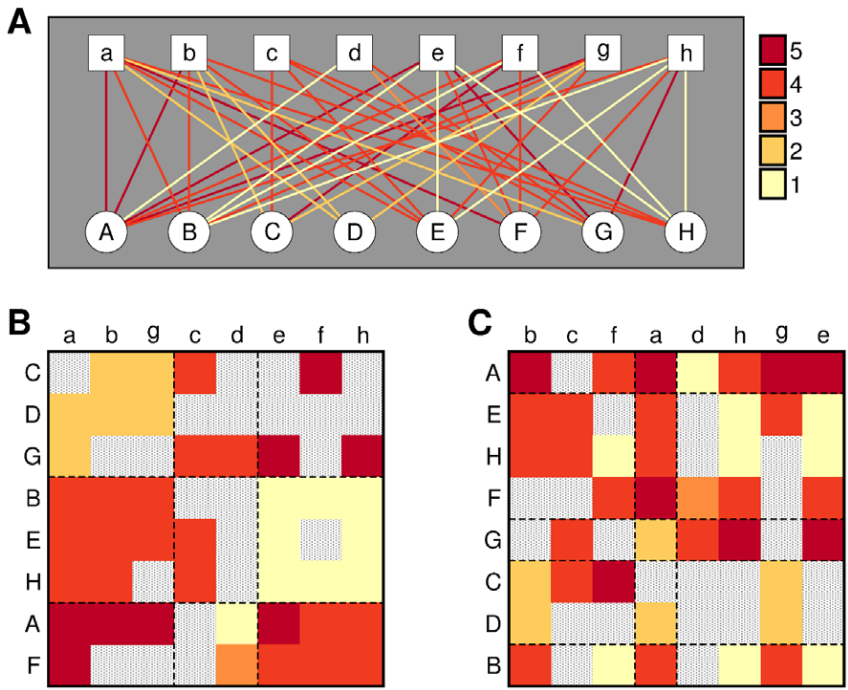

**Figure 1. Predicting preferences using stochastic block models.** (A) Users $A$–$H$ rate movies $a$–$h$ as indicated by the colors of the links. (B-C) Matrix representation of the ratings; patterned gray elements represent unobserved ratings. Different partitions of the nodes into groups (indicated by the dashed lines) provide different explanations for the observed ratings. The partition in (B) has much explanatory power (low $\mathcal{H}$) because ratings in each pair of user-item groups are very homogeneous. For example, it seems plausible that $C$ would rate item $a$ with a 2, given that all users in the $\{C,D,G\}$ group give a 2 to all items in group $\{a,b,g\}$. Conversely, the partition in (C) has very little explanatory power. According to Eq. 4, the predictions of (B) contribute much more than those of (C) to the inference of unobserved ratings.
doi:10.1371/journal.pone.0044620.g001

## Results

### Performance Comparison on Model Ratings

To investigate how our approach performs compared to the benchmark algorithms, and in what situations it works better or worse, we start by generating model dichotomous like/dislike ratings as follows. First, each item $i$ is assigned an intrinsic quality $Q_i \in [0,1]$. Additionally, items and users are partitioned into groups, and each user $u$ has an a priori preference $P(\sigma_u, \sigma_i) \in [0,1]$ for item $i$, where $\sigma_u$ and $\sigma_i$ are the user and item groups, respectively. Then, the probability that $u$ rates $i$ with $r = 1$ ("like", as opposed to $r = 0$, "dislike") is

$$p(r_{ui} = 1) = Q_i^{(1-\alpha)} P(\sigma_u, \sigma_i)^\alpha , \qquad (7)$$

where $\alpha \in [0,1]$ is a parameter that enables us to interpolate between a situation in which the intrinsic quality of the item is the only relevant factor ($\alpha = 0$) and a situation in which a priori preferences are the only relevant factor ($\alpha = 1$).

In Fig. 2, we show the performance of the different algorithms when applied to model ratings. When the intrinsic quality is the dominant factor in user ratings ($\alpha < 0.5$), all algorithms perform similarly well. Of note, in the limiting case where intrinsic quality is the only relevant factor ($\alpha = 0$), the naive recommender is the optimal predictor and does indeed perform slightly better than the others.

Conversely, when a priori preferences start playing a significant role ($\alpha > 0.5$) algorithms start to differ in their performance. As expected, the naive recommender performs poorly in this regime and becomes totally uninformative when $\alpha=1$. The performances of the other algorithms are closer, but SBM is significantly and consistently the most accurate.

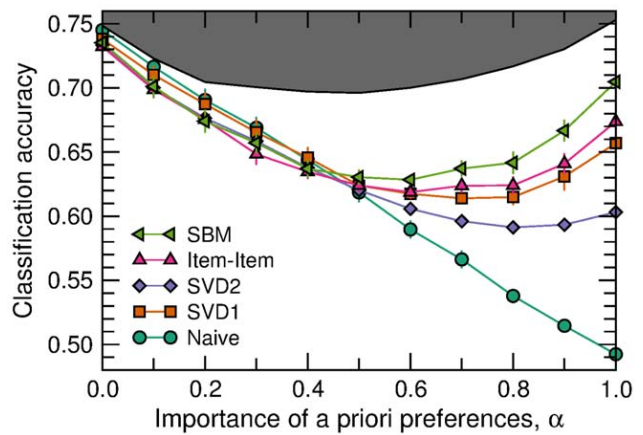

**Figure 2. Algorithm comparison for model ratings.** We show the prediction accuracy (that is, the fraction of correct rating predictions) as a function of the parameter $\alpha$ that measures the importance of a priori preferences as opposed to intrinsic item quality (see text for details). The black line represents the optimal prediction accuracy, which would be obtained if the algorithms were able to estimate exactly the probability of each rating. For all the simulations we use: $n_u = 100$ users organized in 5 groups; $n_i = 100$ items organized in 5 groups; $P(\sigma_u, \sigma_i)$ uniformly distributed in $[0,1]$; 4,000 observed ratings; and 1,000 ratings in the test set.
doi:10.1371/journal.pone.0044620.g002

Of course, for $\alpha=1$ model ratings are generated according to a block model, so the SBM approach is expected to work best. However, it is worth pointing out that at least for these model ratings, the most advanced collaborative filtering approaches are never the most accurate, regardless of the value of $\alpha$–either they perform slightly worse than the naive recommender, or they perform significantly worse than the SBM. Since these collaborative filtering approaches are known to be much more accurate in real data than the naive approach (indeed, they are consistently the most accurate among collaborative filtering methods in the literature [4]), our results on model ratings suggest that the SBM algorithm has the potential to provide good estimates on real data. Additionally, our approach also seems to be the most robust because it never provides estimates that are significantly worse than those produced by any other algorithm.

## Performance Comparison on the MovieLens Dataset

The MovieLens dataset is one of the gold-standards for testing collaborative filtering algorithms [4]. It contains 100,000 real ratings ($r\in\{1,2,3,4,5\}$) from 943 users on 1,682 movies, which were collected through the MovieLens web site (movielens.umn.edu) during the seven-month period from September 19th, 1997 through April 22nd, 1998. For purposes of validation, the dataset is organized in five different splits, each containing a training set $R^O$ with 80,000 ratings and a test set with 20,000 ratings.

As we show in Fig. 3, our algorithm is the most accurate for all and each of the test sets, both in terms of the classification accuracy (that is, the fraction of predictions that are exactly correct) and in terms of the mean absolute error (the mean of the absolute value of the difference between the predicted and the real ratings).

To fully appreciate the importance of our improvement over existing algorithms, it is worth noting that, in terms of classification accuracy, the average improvement of the SBM approach over the best recommender (the Item-Item algorithm) represents a $38\%\pm3\%$ of the improvement of the best recommender over the baseline (naive recommender). The improvement of SBM over SVD, relative to the improvement of SVD over the baseline, is $99\%\pm6\%$. These are major improvements, especially when compared to the differences that could be attributed to implementation details, which are small as shown by the difference in performance between SVD1 and SVD2 (Fig. 3).

## Characteristics of Sampled Partitions

As we have pointed out before, our approach offers the opportunity to study the collections of groupings that have the most explanatory power, namely, those that the Metropolis sampler visits. The MovieLens dataset includes some demographic information about users (such as gender and age) as well as some characteristics of the movies (such as genre). We use this information to assess whether the groupings we sample are indeed correlated with these user and movie characteristics, even when the stochastic block model does not take this information into account.

In particular, we study the co-classification of users [30], that is, the probability that two users belong to the same group

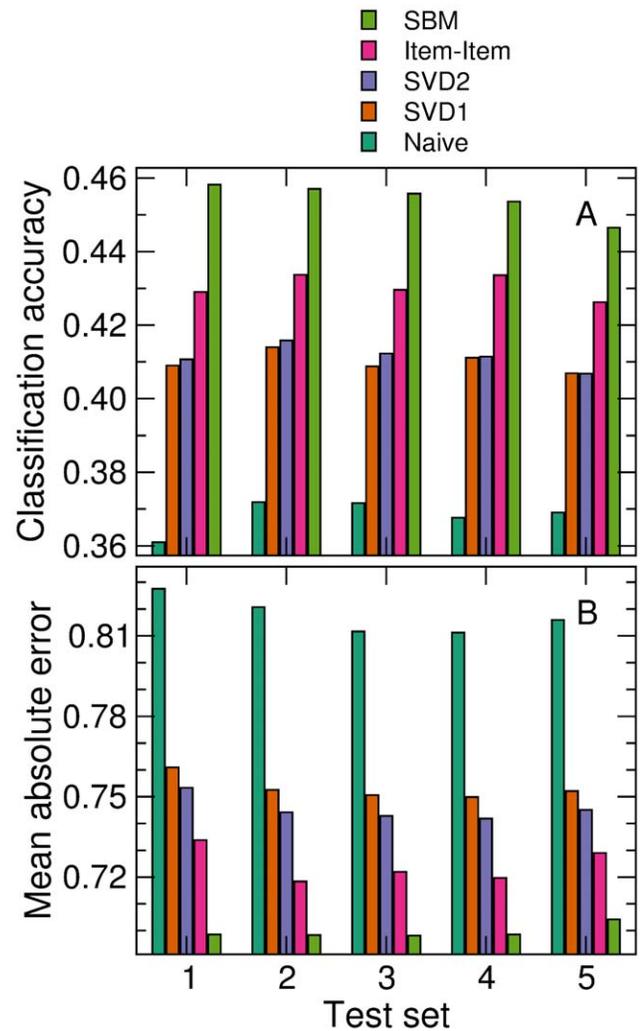

**Figure 3. Algorithm comparison for real ratings from the MovieLens dataset.** Each test set corresponds to a split of the 100,000 ratings in the complete dataset into 80,000 observed ratings and 20,000 test ratings. (A) Classification accuracy is the fraction of 1–5 ratings that are exactly predicted by each algorithm. (B) Mean absolute error is the mean absolute deviation of the prediction from the actual rating.
doi:10.1371/journal.pone.0044620.g003

$$p_{\text{SBM}}(\sigma_{u_1}=\sigma_{u_2}|R^O) = \frac{1}{Z}\sum_{\substack{P_U\in\mathcal{P}_U \\ P_I\in\mathcal{P}_I}} \delta(\sigma_{u_1},\sigma_{u_2})e^{-\mathcal{H}(P_U,P_I)}. \quad (8)$$

We then plot the probability that a pair of users have the same gender and their average age difference as a function of their co-classification probability (Fig. 4A). We observe that user co-classification is strongly correlated with both demographic properties. For example, a pair of users that are very unlikely to belong to the same group have the same gender 58% of the times, whereas pairs of nodes that are almost surely in the same group have the same gender 83% of the times.

Similarly, we plot the genre overlap (see Methods) between two movies as a function of their co-classification probability (Fig. 4B). Again, we observe a strong correlation, which indicates that the stochastic block model correctly picks groups that are related to movie content, even without having access to such information.

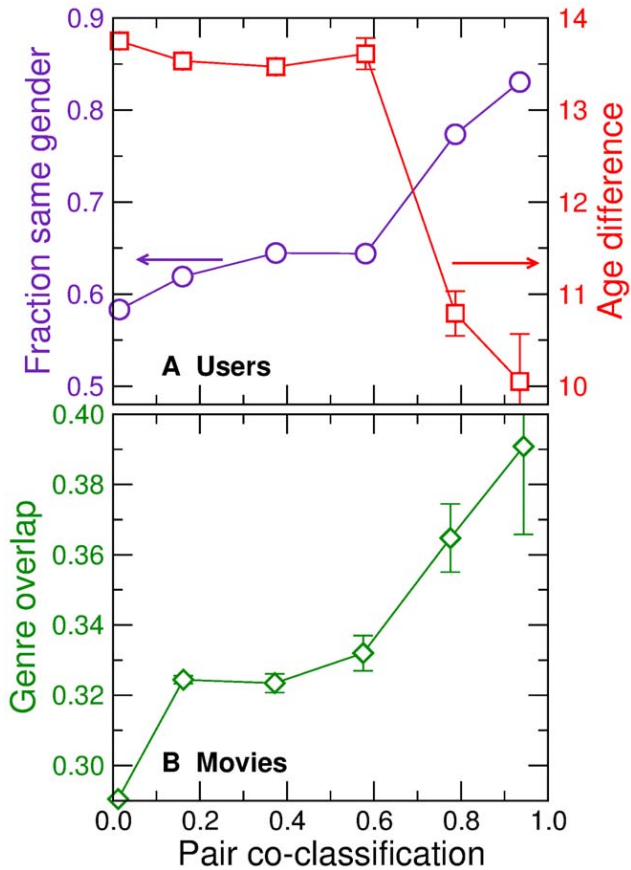

**Figure 4. Characteristics of sampled partitions.** We calculate how often each pair of users (A) or movies (B) are co-classified in the same group in the sampled partitions. (A) Probability that a pair of users have the same gender (circles), and their age difference (squares), as a function of their co-classification frequency. (B) Overlap between the genres of a pair of movies (see Methods) as a function of their co-classification frequency.
doi:10.1371/journal.pone.0044620.g004

## Discussion

We have shown that a Bayesian approach based on the block structure of social networks gives predictions of human preferences that are significantly and considerably more accurate than leading collaborative filtering recommender algorithms.

Like any other approach, ours has shortcomings. In particular, it is worth noting that the gain in accuracy comes at the expense of computational cost–Metropolis sampling of the user and item partition space is computationally demanding. Although we are able to run the algorithm on the MovieLens dataset with approximately 1,000 users and items and 100,000 ratings, handling even one order of magnitude more might be challenging. Besides parallelizing the sampling process (which is straightforward), we think that two approaches could significantly reduce the computational cost: (i) finding analytical approximations to Eq. (4), or even an exact series expansion in terms of the ratings matrix; (ii) implementing a believe propagation algorithm [23,31] to replace Monte Carlo sampling.

In any case, we consider that the advantages of our approach outweigh its shortcomings. Not only does our algorithm provide better predictions, but also has some desirable features: it is mathematically rigorous, it is based on plausible social models, and it sheds light on decision-making processes.

With respect to mathematical rigor, the Bayesian approach is the complete and correct probabilistic treatment of the observations. As a result, we obtain an estimate of the whole probability distribution for each rating $p(r_{ui} = r|R^O)$. From this, we can choose how to make predictions (the most likely rating, the mean, the median, an others). In contrast, recommender systems like those based on matrix factorization give predictions that, in general, are not feasible ratings (for example, $r^*_{ui} = 0.65$ when $r_{ui} \in \{0,1\}$) or that may even be outside the rating range (for example, $r^*_{ui} = 1.2$ when $r_{ui} \in \{0,1\}$). Additionally, these algorithms assume that ratings are linearly spaced in the "psychological scale" of users (that is, that the difference between $r_{ui} = 5$ and $r_{ui} = 4$ is the same as between $r_{ui} = 2$ and $r_{ui} = 1$), which is known not to be true [4].

Finally, our approach is based on models that were originally defined and are widely used to explain how social agents establish relationships, and is therefore in a better position to illuminate which social and psychological factors determine human preferences. As an interesting byproduct of this, we note that it is possible to use our approach to infer demographic properties from ratings alone, a subject that is of much current interest [32].

## Methods

### Derivation of the Rating Equations

Here, we show how we derive the expressions for the probability of a given rating (Eq. (4)) starting from the general Bayesian formulation of the problem (Eq. (2)). In a stochastic block model, users and items are partitioned into groups and the probability that a user rates an item with $r_{ui} = r$ depends, exclusively, on the groups $\sigma_u$ and $\sigma_i$ to which the user and the item belong, that is

$$p(r_{ui} = r|M) = q_r(\sigma_u, \sigma_i) \in [0,1], \quad (9)$$

with $\sum_r q_r(\sigma_u, \sigma_i) = 1$. Other than this normalization constraint, $q_r(\sigma_u, \sigma_i)$ can take any value between 0 and 1.

As in the main text, we consider the case in which ratings can take $K$ different values $r \in \{1, \ldots, K\}$. In this case, a model $M = (P_U, P_I, \{Q_1, \ldots, Q_K\})$ is completely specified by a partition $P_U$ of the users, a partition $P_I$ of the items, and $K$ matrices $Q_r$, $r = 1, \ldots, K$, whose elements are $q_r(\alpha, \beta)$. Then the likelihood of a model is

$$p(R^O|M) = \prod_{\alpha \in P_U} \prod_{\beta \in P_I} \prod_{i=1}^{K} q_i(\alpha, \beta)^{n^i_{\alpha\beta}}, \quad (10)$$

where $n^i_{\alpha\beta}$ is the number of $i$-ratings observed from users in group $\alpha$ to items in group $\beta$.

Putting together Eqs. (2), (9) and (10), and under the assumption of no prior knowledge about the models ($p(M) = const.$), we have

$$p(r_{ui} = r|R^O) = \frac{1}{Z} \sum_{\substack{P_U \in p_U \\ P_I \in p_I}} \int d\mathbf{Q} \, q_r(\sigma_u, \sigma_i) \prod_{\alpha \in P_U} \prod_{\beta \in P_I} \prod_{i=1}^{K} q_i(\alpha, \beta)^{n^i_{\alpha\beta}}; \quad (11)$$

where the integral is over all $q_i(\alpha, \beta)$ within the subspace that satisfies the normalization constraints $\sum_i q_i(\alpha, \beta) = 1$. These integrals factorize and one is left with only two types of integrals to solve. For $\alpha = \sigma_u$, $\beta = \sigma_i$ and $i = r$ we have (without loss of generality we consider the case $r=1$ and, for clarity, we drop the dependence of $q_i$ on $\alpha$ and $\beta$)

$$\int_0^1 dq_1\, q_1^{n^1+1} \int_0^{1-q_1} dq_2\, q_2^{n^2} \cdots \int_0^{1-q_1-\ldots-q_{K-2}} dq_{K-1}\, q_{K-1}^{n^{K-1}}$$
$$(1-q_1-\ldots-q_{K-1})^{n^K}$$
$$= \frac{(n^1+1)!n^2!\ldots n^K!}{(n^1+n^2+\ldots+n^K+K)!} \;.$$

For all other terms we have

$$\int_0^1 dq_1\, q_1^{n^1} \int_0^{1-q_1} dq_2\, q_2^{n^2} \cdots \int_0^{1-q_1-\ldots-q_{K-2}} dq_{K-1}\, q_{K-1}^{n^{K-1}}$$
$$(1-q_1-\ldots-q_{K-1})^{n^K}$$
$$= \frac{n^1!n^2!\ldots n^K!}{(n^1+n^2+\ldots+n^K+K-1)!} \;.$$

Using these expressions in Eq. (11), one obtains Eq. (4).

### Sampling of the Partition Space

Uniformly sampling the space of users' and movies' partitions is necessary to get accurate estimates of $r_{ui}$ (Eq. (4)). The simplest way to sample users (or movies) partitions is by considering a random initial partition and then attempting moves of individual users from their current group to a new group, which is selected uniformly at random. However, this approach has the shortcoming of implicitly considering groups as distinguishable–for example, if node $A$ is alone in group 1 and we move it to an empty group, the partition has not changed but the algorithm considers it as different.

In fact, when there are as many potential user groups as there are users, considering groups as distinguishable has the effect of over-counting partitions by a factor $(N_u-k_u)!/N_u!$, where $N_u$ is the number of users and $k_u$ is the number of non-empty user groups in the partition.

Since, as we have said, sampling over partitions with distinguishable groups is easiest to implement, in practice we use a modified Hamiltonian that "penalizes" partitions that are otherwise over-counted

$$\mathcal{H}'(P_u,P_m) = \mathcal{H}(P_u,P_m)$$
$$- \log[(N_u-k_u)!]$$
$$- \log[(N_m-k_m)!] \;, \quad (12)$$

where $\mathcal{H}(P_u,P_m)$ is given by Eq. (5).

Note that the additional terms in Eq. (12) are *not* a priori penalties to avoid over-fitting by models with many groups, but rather corrections to a sampling process that would otherwise be biased. The over-fitting problem, which is common to other approaches to inference of block models [23], is automatically solved by our marginalization over the $q_r$ probabilities.

For infinitely long samplings of the space of partitions, the correction in Eq. (12) exactly cancels the over-counting of certain partitions that our sampling method causes. For finite sampling times, one cannot be sure that the whole partition space is uniformly sampled. To minimize this potential problem, we run short, parallel and independent sampling processes in different regions of the partition space, as opposed to a single long sampling process. This slightly improves our predictions of model and real ratings (although the improvement is small compared to the difference between our algorithm and other algorithms' performance).

### Benchmark Algorithms

In the naive recommender (Naive), the rating of user $u$ for item $i$ is simply the average rating of $i$ by all users:

$$r_{ui}^N = \frac{\sum_{u' \in U_i} r_{u'i}}{|U_i|}, \quad (13)$$

where $U_i$ is the set of users that rated item $i$ and $|U_i|$ is the number of users in that set.

The matrix factorization method based on singular value decomposition (SVD) works as follows [17]. The matrix of ratings $R$ (with a number of rows $n_u$ that coincides with the number of users, and a number of columns $n_i$ that coincides with the number of items) can be decomposed, using singular value decomposition, into

$$R = P\, Q, \quad (14)$$

where $P$ is a $n_u \times n_i$ matrix and $Q$ is a $n_i \times n_i$ matrix.

If we denote the rows of matrix $P$ as $p_u^T$ and the columns of $Q$ as $q_i$, then individual ratings satisfy $r_{ui} = p_u^T q_i$. For the purpose of making recommendations, it is convenient to pose the decomposition problem as an optimization one; indeed, one can prove that $P$ and $Q$ are the solution of

$$\{p_u, q_i\} = \arg\min_{\tilde{p}_u, \tilde{q}_i} \sum_{u,i} (r_{ui} - \tilde{p}_u^T \tilde{q}_i)^2 \;. \quad (15)$$

In practice, to estimate unobserverd ratings one needs to take into consideration a number of important issues. First, SVD factorization can have a prohibitive computational cost because we typically deal with large $n_u$ and $n_i$, so the problem has to be dimensionally reduced. Second, only some user-item pairs are observed (namely, those $(u,i) \in R^O$). And third, users and items can have rating biases (for example, some users rate items higher than others, and some items are systematically highly rated).

Ultimately, unobserved ratings $r_{ui}^*$ are estimated using

$$r_{ui}^* = p_u^{*T} q_i^* + \mu + b_u + b_i \;, \quad (16)$$

where $b_u$ and $b_i$ are the biases of users and items respectively and $\mu$ is the average rating in $R^O$. The vectors $p_u^*$ and $q_i^*$ are dimensional reductions of the original $p_u$ and $q_i$, and have length $K < n_u, n_i$. They are obtained by solving the optimization problem

$$\{p_u^*, q_i^*\} = \arg\min_{\tilde{p}_u, \tilde{q}_i} \sum_{(u,i) \in R^O} \left(r_{ui} - \tilde{p}_u^T \tilde{q}_i - \mu - b_u - b_i\right)^2$$

$$+ \lambda \sum_{u,i} \left(\|\tilde{p}_u\|^2 + \|\tilde{q}_i\|^2\right) . \qquad (17)$$

As Funks originally proposed [17] we solve this problem numerically using the stochastic gradient descent algorithm [33]. In our implementation of the algorithm (SVD1), we use $K = 200$. In the LensKit implementation of the algorithm (SVD2) we set $K = 50$ and a learning rate of 0.002 as suggested in Ref. [4].

Finally, the algorithm based on the similarity between items (Item-Item) works as follows [29]. One starts by defining a similarity between items, which in our case is the cosine between the item rating vectors (conveniently adjusted to remove user biases towards higher or lower ratings [29]). The predicted rating $r_{ui}$ is the similarity-weighted average of the $K$ closest neighbors of $i$ that user $u$ has rated. Once more, we use the default, optimized implementation of the algorithm in LensKit [4] ($K = 50$).

### Movie Genre Overlap

Each movie $i$ in the MovieLens dataset is labeled with one or more genres $G_i$. We define the genre overlap $o_{i_1, i_2}$ between two movies as the Jaccard index of the corresponding genre sets

$$o_{i_1, i_2} = \frac{|G_{i_1} \cap G_{i_2}|}{|G_{i_1} \cup G_{i_2}|} , \qquad (18)$$

that is, the ratio between the number of genres shared by the two movies and the total number of genres with which they are labelled.

## Author Contributions

Conceived and designed the experiments: RG AL EM MSP. Performed the experiments: RG AL. Analyzed the data: RG AL EM MSP. Contributed reagents/materials/analysis tools: RG AL EM MSP. Wrote the paper: RG AL EM MSP.